# Spatio-Temporal Imaging of the Acoustic Field Emitted by a Single Copper Nanowire

Cyril Jean,[1, *] Laurent Belliard,[1] Thomas W. Cornelius,[2] Olivier Thomas,[2] Yan Pennec,[3] Marco Cassinelli,[4] Maria Eugenia Toimil-Molares,[4] and Bernard Perrin[1]

[1] *Sorbonne Universités, UPMC Université Paris 06, CNRS UMR 7588, Institut des NanoSciences de Paris, F-75005, Paris, France*
[2] *Aix-Marseille Université, CNRS UMR 7334, IM2NP, F-13397 Marseille Cedex, France*
[3] *Institut d'électronique, de microélectronique et de nanotechnologie (IEMN), UMR CNRS 8520, UFR de physique, Université de Lille-1, Cité scientifique, 59652 Villeneuve-d'Ascq cedex, France*
[4] *GSI Helmholtz Centre for Heavy Ion Research, D-64291 Darmstadt, Germany*
(Dated: September 23, 2016)

The monochromatic and geometrically anisotropic acoustic field generated by 400 nm and 120 nm diameter copper nanowires simply dropped on a 10 µm silicon membrane is investigated in transmission using three-dimensional time-resolved femtosecond pump-probe experiments. Two pump-probe time-resolved experiments are carried out at the same time on both side of the silicon substrate. In reflection, the first radial breathing mode of the nanowire is excited and detected. In transmission, the longitudinal and shear waves are observed. The longitudinal signal is followed by a monochromatic component associated with the relaxation of the nanowire's first radial breathing mode. Finite Difference Time Domain (FDTD) simulations are performed and accurately reproduce the diffracted field. A shape anisotropy resulting from the large aspect ratio of the nanowire is detected in the acoustic field. The orientation of the underlying nanowires is thus acoustically deduced.

Coherent acoustic phonons with very high in-depth spatial resolution can be generated through the absorption of a femtosecond laser pulse by a nano-structure, such as a semiconducting or metallic thin film[1] or multiple quantum wells[2]. Such generated acoustic waves exhibit a wavelength in the order of magnitude of tens of nanometers, which is proved useful for non-invasive ultrasonic imaging in biological tissues[3, 4] or for very high frequency optomechanics[5]. Nonlinear effects could even increase this in-depth spatial resolution by decreasing the spatial extent of the acoustic pulse[6]. However, the lateral resolution of such laser ultrasonic measurements is still limited by the diffraction of light that imposes a micrometric lateral resolution. In order to improve the lateral resolution down to the sub-micrometer scale, various approaches have been suggested. For example, near-field optical miscroscopy (NSOM) was incorporated into the picosecond ultrasonics experimental setup [7, 8]. In addition, an opto-acoustic scanning acoustic microscope has also been reported, claiming a lateral resolution below 100 nm[9, 10]. A Stimulated Emission Depletion (STED) inspired approach has also reduced the acoustic spot lateral dimension to 140 nm[11]. The use of nanostructures as acoustic transducers intergrated in scanning acoustic microscopy systems is also promising. Yang et al.[12] reported the use of a nanowire as an acoustic transducer to optically generate and guide an acoustic pulse. This nanowire can be integrated on an AFM cantilever used to sweep the underlying sample. However, although confined acoustic modes in individual metallic nano-objects is a well-studied topic [13–17], the propagation of acoustic waves in nanometric waveguides[18–21] and the characterization of the ability to generate an acoustic wave into the substrate is scarce[22].

In this letter, we propose and demonstrate the use of copper nanowires as efficient nanometric, monochromatic, tunable and anisotropic acoustic transducers. First, investigations of the radial breathing mode of the nanowires of 120 nm and 400 nm diameter are performed using time-resolved pump-probe spectroscopy in the common front side reflectivity setup. Then, a back side transmission setup is used to detect any acoustic feature generated by the nanowire and reaching the other side of the silicon membrane. This approach based on the spatial imaging of the acoustic field enables us to highlight the shape anisotropy of the nanowire-generated acoustic field. Finite Difference Time Domain (FDTD) simulations are compared to the experimental acoustic field data and may allow a rough estimation of the nanowire-substrate contact size. The imaging of the transmitted 8 GHz and 27 GHz monochromatic acoustic waves through a 10 µm silicon membrane is a very important step towards the use of nano-objects as acoustic transducers for acoustic imaging with nanometric lateral resolution.

Because of their high aspect ratio, their quasi-perfect cylindricity and their smooth contour surface, we have chosen copper nanowires grown by electrodeposition in polycarbonate etched ion-track membranes[23] as model systems to study the confined[24] and propagating[18] acoustic phonons in single nano-objects. The details of their fabrication are described in the supplementary information. Here, these copper nanowires are considered as monochromatic acoustic transducers when simply in contact with a silicon substrate. Indeed, as previously shown[24–26], the huge increase in the quality factor of a nanowire's radial breathing mode when suspended in air is interpreted as the suppression of the dissipation channel through the nanowire-substrate contact. In this work, we go one step further, and are able to detect the dissipated acoustic energy in transmission after its propagation through the substrate. We investigate two types of nanowires with length $L > 10$ µm and diameter $d = 120$ and $d = 400$ nm (Figure 1a). According to the continuum mechanics model that holds at these sizes[27, 28], the expected frequencies $f$ of our mechanical nanoresonators are in the range $f \sim c_L/d \sim 10$ GHz to 30 GHz. At room temperature, these high frequency acoustic phonons suffer severe attenuation in silicon. Consequently the experiments are performed on 10 µm thick silicon membranes obtained by a wet etching in KOH on a 2 inches silicon on insulator (SOI) wafer (Figure 1b,c and d). Experimental details on the fabrication are provided in the supplementary information (Figure S3).

---

* cyril.jean@insp.upmc.fr

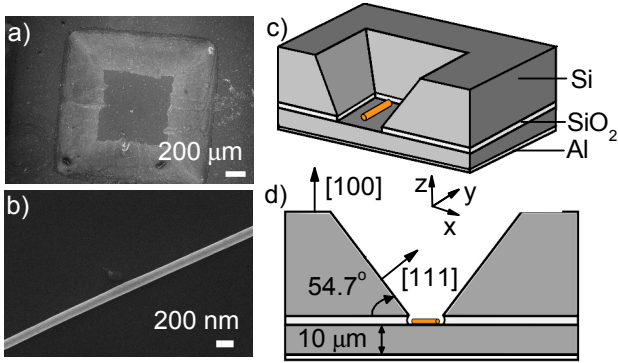
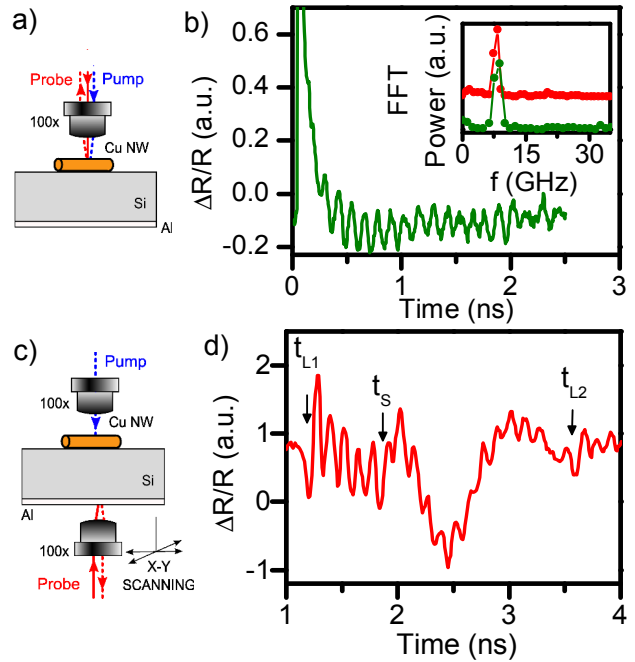

FIG. 1. (a) Close SEM view of a 120 nm copper nanowire supported by the silicon substrate. (b) SEM top view of the etched silicon on insulator wafer. (c) 3D schematic of the etched 100 µm SOI wafer, resulting in a 10 µm silicon membrane with a squared surface topped with a thin 100 nm aluminium layer. (d) Side schematic view of a nanowire resting at the bottom of a silicon membrane.

FIG. 2. (a) Front side configuration. Both the blue pump laser and the red probe laser are focused on a single 400 nm diameter copper nanowire using the same 100× objective with a numerical aperture NA = 0.95. (b) Relative reflectivity $\Delta R/R$ obtained in the front side configuration Figure 1a. It exhibits an oscillatory signal with a period $T = 125$ ps. Inset : normalized FFT power spectrum of the derived signal of Figure 2d (in red) and Figure2b (in green). (c) Back side configuration. The two laser beams are focused on different sides of the silicon membrane. The probe is mounted on a piezoelectric stage in order to map the displacement field on a maximum 150 µm × 150 µm area. (d) Relative reflectivity $\Delta R/R$ obtained in the back side configuration Figure 1c. It exhibits an oscillatory component with a period $T = 125$ ps.

The elastic response of the 400 nm diameter copper nanowires is investigated using a femtosecond optical pump-probe technique described extensively in previous works[29] and in the supplementary information. The mechanical vibrations of the copper nanowires are first studied using the common front side setup where both the pump and the probe laser beams are colinear and focused using the same objective (Figure 2a and S4). Using preliminary SEM experiments (Figure 1b), a single nanowire is located and the subsequent single particle experiment is performed. In this approach, a first blue pump pulse excites the conduction electrons of the metallic nano-object. The subsequent electron cooling by electron-lattice interactions[30] results in the heating of the crystal lattice finally launching the dilatation of the nanowire[31, 32]. Because of the electrons fast diffusion, this ultrafast heating is not confined to the surface of the nano-object, the whole width of the nanowire subsequently undergoes periodic radial contractions and expansions. These isotropic volume oscillations correspond to the fundamental breathing mode of the nanowire[24, 26]. These periodic lattice oscillations induce a modulation of the metal refractive index, which shows up by a periodic variation of the relative probe reflectivity $\Delta R/R$ (Figure 2b). The oscillatory signal is superimposed to an exponentially decaying background as the nanowire cools down by transferring its energy to the substrate. The oscillation frequency $f_0 = 8.0 \pm 0.2$ GHz is extracted from the time trace by either a fast Fourier transform (FFT) or a numerical fitting with a damped sinusoidal function superimposed to an exponentially decaying background. Both methods lead to a quality factor $Q$ on the order of 20. This frequency is compared to the theoretical first radial breathing mode frequency of a free, elastically isotropic and infinite copper cylinder, $f_0 = \phi_0 v_L/(2\pi a)$, where $a = 200 \pm 15$ nm is the radius of the nanowire and $\phi_0$ is the first root of the equation (1)

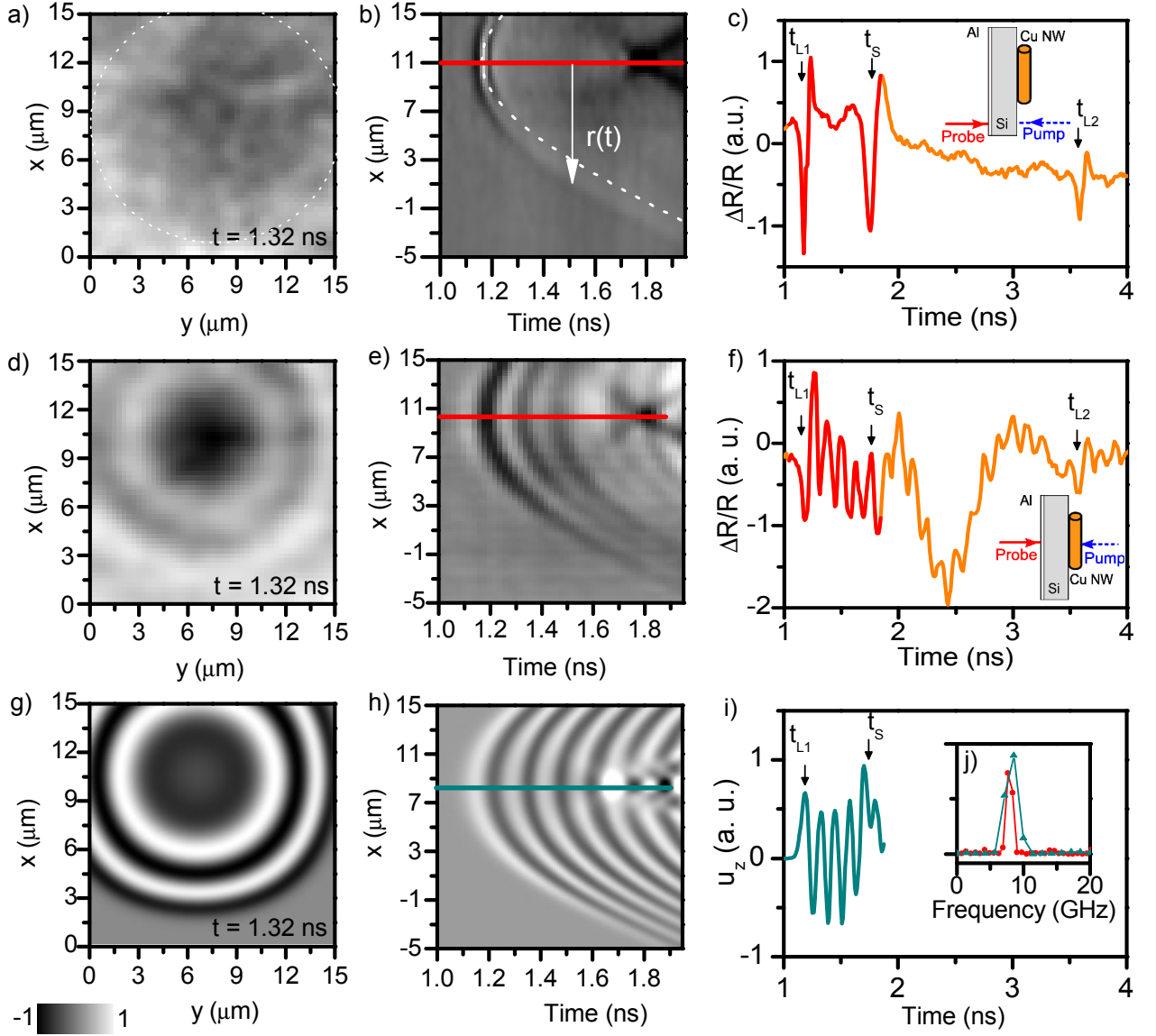

FIG. 3. (a), (b), (c) Reference mapping : the pump hits directly the silicon membrane (schematic inset Figure 3c). (d), (e), (f) The pump excites the first radial breathing mode of the 400 nm diameter copper nanowire (schematic inset Figure 3f) which generates a monochromatic longitudinal acoustic wave leading to concentric rings. (g), (h), (i) FDTD simulation : the acoustic source is a 8 GHz monochromatic longitudinal displacement imposed on the surface of the silicon layer. (j) FFT power spectrum of the red part of the Figure 3f and the turquoise part of the Figure 3i. (a), (d), (g) Two-dimensional $x-y$ surface displacement mapping at $t_{\mathrm{map}} = 1.32$ ns. (b), (e), (h) Two-dimensional $x-t$ surface displacement mapping at fixed $y_0 = 7\,\mu$m. (c), (f), (i) One-dimensional (time domain) surface displacement recording on the position $x_0 = 10\,\mu$m and $y_0 = 7\,\mu$m. The red (or turquoise) part of the signal matches the corresponding red (or turquoise) line on the Figure 3b, e and h.

$$2\left(\frac{v_T}{v_L}\right)^2 J_1(\phi_n) = \phi_n J_0(\phi_n) \quad (1)$$

where $v_L = 4.5 \times 10^3$ m·s$^{-1}$ and $v_T = 2.2 \times 10^3$ m·s$^{-1}$ are respectively the longitudinal and transversal speed of sound in bulk polycrystalline copper[33]. $J_0$ and $J_1$ are respectively the Bessel functions of the first kind of order 0 and 1. The theoretical breathing mode frequency reaches $f_0 = 8.0 \pm 0.3$ GHz, where the frequency uncertainty is dictated by the 15 nm fluctuations in the nanowire diameter. Owing to the radial displacement of this breathing mode, the acoustic energy of the nanowire is expected to leak towards the substrate mainly as a longitudinal wave. By flipping a mirror, the experimental setup is thus switched to the backside configuration (Figure 2c and S4). The nanowire remains excited by the blue pump laser, whereas the red probe laser is now focused on the other side of the silicon membrane. The relative reflectivity trace recorded using a Michelson interferometer is plotted in Figure 2d. The origin of the plot corresponds to the electronic coincidence in the frontside configuration. The longitudinal deformation at $t_{L1} = 1.18 \pm 0.02$ ns is followed by a monochromatic oscillation at the same frequency $f_0 = 8.0 \pm 0.4$ GHz as observed in the frontside configuration (inset Figure 2b). As expected, these oscillations present a quality factor close to that observed in the frontside configuration even if an accurate comparison is not reachable because the damping time of the oscillations is on the order of magnitude of the time delay between the two longitudinal echoes. The second longitudinal echo appears at $t_{L2} = 3.56 \pm 0.02$ ns after an additional round trip. This second echo is also followed by oscillations. Owing to the well known longitudinal speed of sound along the $<100>$ direction of silicon, $v_L^{SI} = 8.43 \times 10^3$ m·s$^{-1}$, one gets the thickness $h = (t_{L2} - t_{L1}) v_L^{SI}/2 = 10.0 \pm 0.2$ μm of the silicon membrane. The shear wave is thus expected at a time delay $t_S = h/v_T^{SI} = 1.84 \pm 0.02$ ns, where $v_T^{SI} = 5.84 \times 10^3$ m·s$^{-1}$ is the transverse speed of sound along the $<100>$ direction of silicon. As the shear wave reaches the surface while the 8.0 GHz longitudinal oscillations are still occuring, the exact time of arrival cannot be accurately determined. But the main reflectivity change produced by the shear wave appears between 1.8 ns and 2.8 ns. This wide shear wave signature will be discussed later on. The perfect agreement between the resonance frequency of the nanowires' radial breathing mode and the frequency of the generated monochromatic longitudinal wave across the membrane is strong evidence that the nanowire acts as a nanoacoustic wave transducer. We noticed this behavior on several different nanowires : whenever the front side configuration signal shows an oscillatory component, the reported emission phenomenon is observed.

A control experiment is additionally performed. Starting with the time-resolved interferometric measurement in Figure 2d (reproduced in the Figure 3f) when the blue pump beam hits the copper nanowire, the sample is smoothly moved laterally so as the pump beam does not excite the nanowire anymore (Figure 3c). In this configuration, the monochromatic oscillation disappears and the three remaining dipolar peaks at $t_{L1}$, $t_{L2}$ and $t_S$ correspond to the longitudinal (direct and after one round trip) and transverse echoes. Both longitudinal and shear waves are effectively excited in the (100) oriented silicon membrane due to the small size of the pump and probe beams[34, 35]. The amplitude of the acoustic echoes is similar between the two experiments but the pump power is 5 mW in Figure 3c and 0.5 mW in Figure 3f. The metallic nanowire is thus accountable for the monochromatic oscillations observed in Figure 3f. The surface displacement produced by the acoustic wave was recorded in time and space by displacing the probe laser which is mounted on a piezoelectric stage allowing a movement of 150 μm with an accuracy of 5 nm. Complementary to the experiments described previously (see Figures 3c and f), two different arrangements of the backside configuration were realized where the pump laser beam excites either the nanowire (Figures 3d,e) or the silicon substrate directly (Figures 3a,b). The first experiment is a two dimensional $x-y$ surface displacement mapping at a fixed time delay $t_{map} = 1.32$ ns after the electronic coincidence in the frontside configuration. Considering the micrometric penetration depth of the light in the silicon, the generation of acoustic waves directly in the silicon substrate is inefficient, explaining the extremely poor contrast in the 2D-spatial mapping shown in Figure 3a (the dotted circle is a guide for the eye). On the contrary, when the acoustic wave is generated by a laser-excited copper nanowire, the generation of concentric acoustic waves is highlighted in Figure 3d. These experiments can be performed at different time delays in order to obtain a movie (Movie S1 and S2). As the patterns possess a quasi circular symmetry, another experiment is performed to obtain informations on how the patterns change as the acoustic waves evolve in time. This second experiment is a two-dimensional $x-t$ surface displacement mapping at a fixed position $y_0 = 7$ μm. The alternation of white and black stripes in Figure 3e which is missing on the reference Figure 3c is characteristic of the longitudinal monochromatic acoustic wave emitted by the copper nanowire. In addition, using the rough isotropic approximation, the longitudinal branches in the Figure3 b and e should follow the equation $r(t) = \sqrt{(v_L(t - t_{L1}) + h)^2 - h^2}$ (dotted line in the Figure 3b). This approximation underestimates the experimental $r(t)$. To investigate the nanowire-mediated generation of acoustic waves beyond the rough isotropic estimation, anisotropic FDTD experiments are calculated using a Fortran code developped at the IEMN in Lille. In order to design the geometry of the acoustic source in this simulation, an estimation of the spatial dimensions of the acoustic source generated by the nanowire deposited on the silicon frontside is needed. For now, the nanowires' excitation is produced by a blue light pulse ($\lambda = 400$ nm) using a 100× objective with a numerical aperture NA = 0.95. The beam diameter at $1/e^2$ is thus measured to be slightly smaller than 1 μm. The excited surface of the silicon substrate may be estimated as the surface of contact between the excited area of the 400 nm nanowire and the silicon. Here, the acoustic source along the nanowire axis is confined to the irradiated part. Possible enlargement associated to guided acoustic phonons[18] along the axis could be neglected due to the huge elastic coupling with the substrate. In a first rough approximation, this source is described in the FDTD simulation by imposing a boundary speed equation

$v_{zz}(x, y, z, t) = \sin(2\pi f_0 t)$
$\qquad \exp\left(8\left((y - y_0)^2/\sigma_y^2 + (z - z_0)^2/\sigma_z^2 + x^2/\sigma_x^2\right)\right)$

where $\sigma_x = 1$ μm is related to the pump beam diameter, $\sigma_y = 0.2$ μm is given by the nanowire diameter and

$\sigma_z = 0.1\,\mu m$ is chosen to be small but nonzero to avoid abrupt discontinuity. The out of plane surface displacement $u_z$ is thus recorded and plotted in Figures 3g-i using the exact same representation as in Figures 3d-f. The time of flight of the longitudinal and transverse echoes (Figure 3i), as well as the concentric rings (Figure 3g) or alternating white and black stripes (Figure 3h) are in good agreement with the experimental data in figures 3d-f. Modelling the copper nanowire as a submicronic monochromatic longitudinal acoustic transducer seems to effectively describe our experimental observations, even though the transverse acoustic wave signature cannot be grasped perfectly by these simulations.

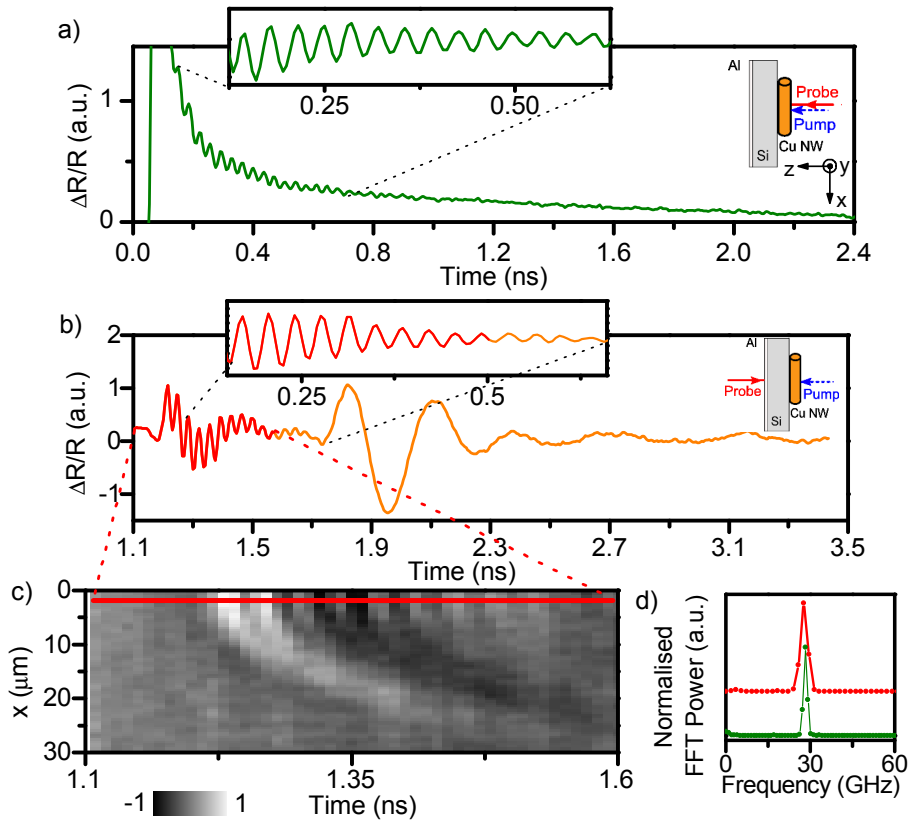

FIG. 4. (a) Relative reflectivity $\Delta R/R$ obtained in the front side configuration (See schematic in the inset). Both the blue pump and the red probe are focused on a single 120 nm diameter copper nanowire using the same 100× objective with a numerical aperture NA = 0.95. The other inset is the derived signal exhibiting a 35 ps oscillatory signal. (b) Relative reflectivity $\Delta R/R$ obtained in the back side configuration (see schematic in the inset) when the blue pump is focused on a single 120 nm diameter copper nanowire. The other inset is the derived signal exhibiting an oscillatory signal with a periodicity $T = 38$ ps (c) Two-dimensional (spatiotemporal) surface displacement mapping along the spatial direction $x$. The red line corresponds to the red part of the oscillatory signal in Figure 4b. (d) Normalised FFT power spectrum of the inset of Figure 4b (in red) and Figure 4a (in green).

According to the simultaneous experiments in the frontside and the backside configurations of Figure 2, the generated acoustic wave is due to the laser-excited radial breathing mode of the copper nanowire. As the first radial breathing mode frequency, $f_0$ is inversely proportional to the radius $a$ of the nanowire, it is expected that using smaller nanowires will result in a higher frequency acoustic transducer. Figure 4 shows the results obtained using copper nanowires with $120 \pm 10$ nm diameter. For these thinner nanowires, the radial breathing mode frequency amounts $f_0 = 26.5 \pm 0.2$ GHz (Figure 4d), which corresponds to a temporal period $T_0 = 38 \pm 3$ ps. Similar to Figure 2, both reflectivity frontside (Figure 4a) and interferometric backside measurements (Figure 4b) are shown. Once again, the monochromatic oscillations obtained using the frontside optical setup (Figure 4 relative reflectivity time trace a) or the backside optical setup (Figure 4 relative reflectivity time trace b) appear to be at the same frequency $f_0 = 26.5 \pm 0.2$ GHz (Figure 4d). Another oscillatory component at ∼ 3 GHz starts around 1.8 ns, at the time delay where the shear wave is expected. This low frequency modulation was also present in Figure 2 at the same time delay but at a lower frequency around ∼ 1.5 GHz. This could be the signature of a low frequency shear acoustic mode generated by the nanowire as the frequency depends on the nanowire's di-ameter. However, as this acoustic feature is missing in the frontside experiment, we are not able to attribute this behavior to any specific acoustic mode. It could be related to some contact mode in the same manner that was previously observed on single gold nanoparticles[36]. Thus, further investigations are still needed to address this question. The insets in these two sub-figures are the derived signals to get rid of the low frequency features in the relative reflectivity signals. Unlike the 8 GHz case, these insets permit a direct and convincing comparison of the damping time of the oscillations between the two cases as the oscillation lifetime is shorter than the time delay between the two longitudinal echoes. Thus, this damping time can be accurately determined and reaches $\tau = 0.2$ ns in transmission to be compared to $\tau = 0.3$ ns in the frontside configuration, which corresponds to a quality factor $Q = \pi f_0 \tau \sim 30$ in good adequation with previous studies on supported metallic nanowires[24, 37]. A pseudocolor $x-t$ spatiotemporal surface displacement mapping along the spatial direction $x$ is presented in the Figure 4c. The white and black stripe alternance, which corresponds to the nanowire breathing mode leaking through the substrate, is only distinguishable directly in front of the generated acoustic wave. As the probe is moved along the $x$ axis, this acoustic feature disappears under the noise. The square root shape of the branches is only visible for the first (and more intense) oscillation. As the wavelength of the 26.5 GHz acoustic wave reaches $\sim 0.3\,\mu$m, a spatial imaging (here along the direction $x$) becomes highly challenging as the probe is focused on a $\sim 1\,\mu$m diameter spot. The attenuation that increases with the square of the frequency is also a critical parameter preventing us to obtain nice spatial mapping images.

In the last part of this article, we focus on the 400 nm diameter nanowires vibrating at 8 GHz. As previously described, the blue pump pulse is focused on a nanowire using a 100× objective with a numerical aperture NA = 0.95. The beam diameter at $1/e^2$ is thus estimated to be under 1 μm. Consequently, the contact surface between an excited 400 nm diameter copper nanowire and the silicon substrate is estimated as a 200 nm × 1 μm rectangle. This elongated acoustic source should produce a geometrically anisotropic emission, due to a smaller size compared to the acoustic wavelength emitted in the silicon. Thus, the orientation of the underlying nanowire should be detected using such an anisotropic acoustic field. However, the $x-y$ surface displacement mapping of the acoustic field produced by a 400 nm diameter nanowire in Figure 3d does not seem anisotropic : the inner ring seems circular and the color along its perimeter is constant and modulated by a strong noise. The FDTD simulation in Figure 3g does not reveal any obvious anisotropy in the acoustic field either. To understand this observation, let us remind the conditions to observe a diffraction phenomenon. Generally speaking, this phenomenom occur if the size of the obstacle or the slit encountered by the wave is close to the spatial wavelength of the wave. In the present case, the upper limit dimensions of the acous-tic source is a 200 nm × 1 µm rectangle. In comparison, the wavelength of a longitudinal acoustic wave at 8 GHz is $\lambda_0 = v_L/f_0 \simeq 1\,\mu m$. Both dimensions of the acoustic source are thus below the caracteristic length of the acoustic wave. A significant diffraction is therefore expected in both directions. The remote position of the observer is a second necessary condition to observe a far field diffracted patern. In our experiments, the diffraction will be significantly stronger if the two dimensional $x-y$ surface displacement mapping is undergone at a larger scale than the 15 µm × 15 µm of Figure 3. To do so, a 27 µm × 27 µm $x-y$ surface displacement mapping is observed at a time delay $t_d$ chosen to be 700 ps after the first longitudinal echo detection in the center of the images (Figures 5a, c and e). In order to avoid any confusion with an elastic anisotropy resulting from the silicon substrate, three different nanowires are investigated. A blue line corresponding to the effective orientation of the long axis of the nanowire is added in the graphs. The orientation of the nanowire is checked using a CCD camera positioned on the probe beam path of the frontside setup. The effective position of the nanowires according to the recorded images is thus confirmed without any ambiguity. If the first one (Figure 5a) is arbitrarily referred to as horizontal (0°), the second one is at 45° and the third one is at −55°. The three corresponding pseudocolor plots of the 27 µm × 27 µm surface displacement mapping are prensented in the Figure 5a,c and e. A substantial modification is also made on the optical setup. Instead of using a 100× objective with a numerical aperture NA = 0.95 to focus the blue pump laser beam, a 50× objective with a numerical aperture NA = 0.5 is preferred. The beam diameter at $1/e^2$ is now estimated to be smaller than 2 µm. The upper bound of the contact surface between the excited part of the nanowire and the silicon membrane is therefore a 200 nm × 2 µm rectangle. Only one dimension of this rectangular source is now smaller than the wavelength $\lambda_0$ of the considered acoustic wave. This dimension should be favorable to the observation of the diffraction. On the contrary, regarding the 2 µm direction of the source, the phenomenon should be missing. While in Figure 3d when the laser beam diameter is slightly under 1 µm we observe concentric rings, the images in Figures 5a, c and e exhibit incomplete rings. Being confident in the nanowires' orientation, the contrast between each ring appears to be maximal in the direction symbolized by an orange line and minimal in the orthogonal direction symbolized by the blue line corresponding to the long axis of the nanowire. These observations are in good agreement with the fact that diffraction is far more efficient in the direction perpendicular to the nanowire. Moreover, the two profile plots related to each image are depicted on the right hand side of the Figure 5. The spatial period between each oscillation decreases as the distance from the epicenter increases. Using the same isotropic approximation as above, it can be shown that the the spatial separation $\Delta R$ between two successive rings is nonlinear (see the supplementary information for details). In the case investigated here, the function decreases from 2.4 to 1.5 µm from the inner to the outer rings. However, the radius of the rings are underestimated (3.9 µm, 6.2 µm, 8.0 µm and 9.6 µm in comparison to the experimental values 6.0 ± 0.2 µm, 8.7 ± 0.3 µm, 10.1 ± 0.3 µm and 11.7 ± 0.3 µm). If this isotropic estimation grasps the good tendency, an anisotropic model is required for quantitative analysis. The same FDTD simulations are performed, modifying only $\sigma_x$ to $\sigma_x = 2\,\mu m$ compared to Figure 3g in the above study. The simulated pseudocolor plot and the corresponding profiles are displayed in Figure 5g and h, respectively. The calculated radiuses of the different rings amount to 6.0 µm, 8.5 µm, 10.3 µm and 11.7 µm, in excellent agreement with the experimental radiuses. The spatial separation effectively decreases between two successive rings from 2.5 µm, to 1.8 µm and finally to 1.4 µm. The FDTD simulations clearly evidence the elastic anisotropy by the fact the concentric rings structures are not perfectly circular. It has to be differentiated from the shape anisotropy resulting from the elongated source. As a last physical insight, the shape anisotropy is quantified *in silico*. The simulation is undergone with $\sigma_y$ varying from 100 nm to 2 µm. A plot similar to Figure 5h is obtained and the ratio between the oscillation amplitude along the orange line and the blue line is extracted. As this ratio can be in principle extracted from experimental images, it could be a possible way to estimate $\sigma_y$ and thus to give some insight on the surface contact between the nanowire and substrate which is a challenging field of study[38]. However, below 1 µm, a variation of the surface of adhesion of ~ 100 nm leads to a relative variation of the ratio inferior to 1% which is not distinguishable owing to the level of noise in our experiments.

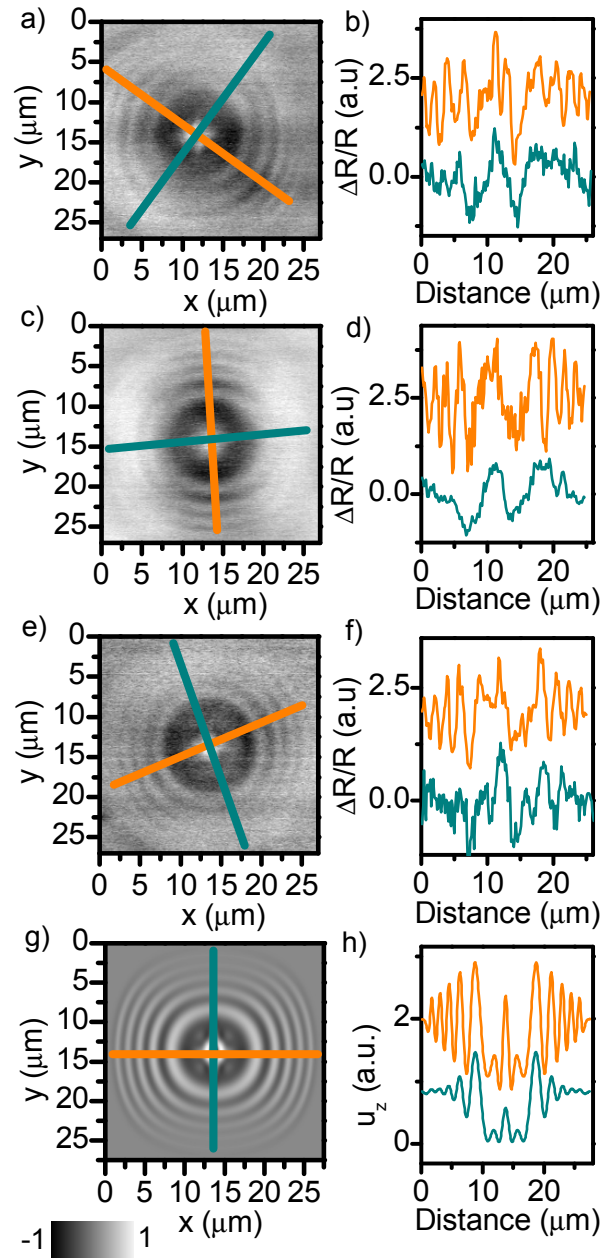

FIG. 5. (a),(c) and (e) Two-dimensional $x-y$ surface displacement mapping 700 ps after the longitudinal echo detection on the central position with a ×50 objective. The three different underlying nanowires are oriented along the blue cyan line. (g) FDTD two-dimensional $x-y$ surface displacement ($u_z$) mapping 700 ps after the longitudinal echo detection on the central position. The acoustic source is $\sim 2\,\mu$m wide along the blue cyan line and $\sim 0.2\,\mu$m wide along the orange line. (b), (d), (f) and (h) Surface displacement profile, respectively along the orange and the cyan line. The offset is for readability purpose.

To conclude, 400 nm and 120 nm diameter electrochemically grown copper nanowires have been used to generate 8 to 27 GHz, monochromatic and longitudinal acoustic phonons with sub-micronic lateral dimensions. Following a standard time-resolved pump-probe spectroscopy experiment in reflectivity to assign the acoustic frequencies to the radial breathing mode of the nanowires, picosecond laser ultrasonic experiments have been performed in transmission on silicon membranes. The as-generated acoustic waves produced by the vibrating nanowires have been imaged in transmission in both time and space domains. The transmitted acoustic field exhibits a strong geometric anisotropy due to the high aspect ratio of the nanowires. The propagation of these gigahertz acoustic phonons generated by the copper nanowires has been modeled using FDTD simulations which gives a good agreement with the experimental data. The transmission of the acoustic waves of the nanowire through a 10 µm silicon substrate definitely shows the good adhesion of nanometric objects simply dropped on their substrate, producing a strong dissipative channel for the nanowire mechanical oscillator. Furthermore, this technique may be used to quantify the surface contact between nanowires and silicon substrates if stronger signal to noise ratios are achieved. Further investigations could also imply the modification of the substrate-wire coupling in order to exalt or inhibit the reported tunable emission of acoustic waves by a nanowire. This demonstration of nanometric longitudinal wave acoustic transducers could also be useful for improved lateral resolution of acoustic microscopy using for instance synthetic aperture focusing techniques[39]. The implementation of such a nanometric transducer on a sweeping device is now needed to pave the way to acoustic microscopy with nanometric lateral resolution.

### ACKNOWLEDGMENTS

MET and MC thank the Deutsche Forschungsgemeinschaft (DFG) for financial support within the priority program SPP 1386.